\documentclass[prc,showpacs,showkeys,superscriptaddress,nofootinbib,twocolumn,floatfix]{revtex4}
\usepackage{amsfonts}
\usepackage{graphicx,color,amsmath,amssymb}

\def\blfootnote{\xdef\@thefnmark{}\@footnotetext}
\newcommand{\eg}{\textit{e.g.}}

\newcommand{\ie}{\textit{i.e.}}

\begin{document}

\title{Modifications of Heavy-Flavor Spectra in $\sqrt{s_{\rm NN}}=62.4~{\rm GeV}$ Au-Au Collisions}
\author{Min He}
\affiliation{Department of Applied Physics, Nanjing University of Science and Technology, Nanjing 210094, China}
\author{Rainer J.\ Fries}
\affiliation{Cyclotron Institute and Department of Physics and Astronomy, Texas A\&M University, College Station, TX 77843, USA}
\author{Ralf Rapp}
\affiliation{Cyclotron Institute and Department of Physics and Astronomy, Texas A\&M University, College Station, TX 77843, USA}

\date{\today}

\begin{abstract}
We calculate open heavy-flavor (HF) production in Au+Au collisions at
$\sqrt{s_{\rm NN}}$=62.4\,GeV utilizing a nonperturbative transport approach
as previously applied in nuclear collisions at top RHIC and LHC energies. The
effects of hot QCD matter are treated in a strong-coupling framework, by
implementing heavy-quark diffusion, hadronization and heavy-flavor meson
diffusion within a hydrodynamic background evolution. Since in our approach
the heavy-flavor coupling to the medium is strongest in the pseudo-critical
region, it is of interest to test its consequences at lower collision energies
where the sensitivity to this region should be enhanced relative to the hotter
(early) fireball temperatures reached at top RHIC and LHC energies. We find
that the suppression and flow pattern of the non-photonic electrons from
heavy-flavor decays at 62.4\,GeV emerges from an intricate interplay of
thermalization and initial-state effects, in particular a Cronin enhancement
which is known to become more pronounced toward lower collision energies.
\end{abstract}


\maketitle

\section{Introduction}
\label{sec_intro}
Heavy quarks (charm and bottom), produced in primordial hard processes in
ultrarelativistic heavy-ion collisions (URHICs), experience the entire
evolution of the hot fireball formed in these reactions. At high transverse
momenta, $p_T\gg m_Q$ (where $m_Q$ is the heavy-quark (HQ) mass), the
quenching and subsequent fragmentation of heavy-flavor (HF) jets allows
one to study the mass and color-charge dependence of parton energy
loss~\cite{Armesto:2005iq}. At low and intermediate $p_T$, however, gluon
radiation is suppressed~\cite{Dokshitzer:2001zm}, and heavy quarks become
unique ``Brownian markers" of elastic thermalization and diffusion processes
in the QCD medium, see Ref.~\cite{Rapp:2009my} for a review. The observation
of a remarkable suppression and elliptic flow of HF electrons at top RHIC
energy~\cite{Abelev:2006db,Adare:2006nq,Adare:2010de} and of HF mesons at the
LHC~\cite{ALICE:2012ab,Abelev:2013lca,Sakai:2013ata,Abelev:2014ipa,Chatrchyan:2012np,CMS:2012wba}
imply a substantial HF coupling to the medium, being dragged by its collective
flow. Different transport models based on perturbative or nonperturbative
interactions have been developed to understand these phenomena at
RHIC~\cite{vanHees:2005wb,Moore:2004tg,Zhang:2005ni,vanHees:2007me,Gossiaux:2009mk,Akamatsu:2008ge,Mazumder:2011nj,Uphoff:2011ad,Alberico:2011zy,He:2011qa,He:2012df}
and
LHC~\cite{Aichelin:2012ww,Uphoff:2012gb,He:2012xz,Lang:2012cx,Alberico:2013bza,Cao:2013ita,He:2014cla}.

A key question in these investigations is whether the HF coupling to the medium
is primarily driven by an increasing temperature (or energy density), or by an
increase in coupling strength in the pseudo-critical region of the
chiral/deconfinement transition. It is therefore of interest to study how the
HF collectivity develops as the collision energy is lowered from top RHIC
energy, to reduce the initial temperatures while still encompassing the
transition region. In the HF sector, this has recently been realized in the
62.4\,GeV run at RHIC~\cite{Mustafa:2012jh,Adare:2014rly}. However, it is known
from light-hadron spectra that initial-state modifications become increasingly
pronounced at lower energies through the Cronin effect, \ie, a nuclear
broadening of the initial spectra due to prescattering prior to the hard
process. This is important in the quantitative interpretation of, \eg, $\pi^0$
suppression at 62.4\,GeV, where the nuclear modification factor, $R_{AA}$,
is significantly less suppressed than at 200\,GeV~\cite{Adare:2012uk}.
For HF observables a significant Cronin enhancement has been observed in
single-electron decay spectra in 200\,GeV d+Au collisions~\cite{Adare:2012yxa},
which is expected to become larger at lower energies.

The aim of the present paper is to quantify the thermalization effect on HF
production in the hot medium at low and intermediate $p_T$ in Au+Au
collisions at $\sqrt{s_{\rm NN}}$=62.4\,GeV. To this end, we employ our
previously developed nonperturbative transport model~\cite{He:2011qa}, with
medium evolution and initial conditions adapted to this energy and
incorporating the Cronin effect in a systematic manner. Specifically, we
calculate the $R_{\rm AA}$ and elliptic flow ($v_2$) of HF electrons and
compare the results to recent PHENIX and STAR data. Since our nonperturbative
HF diffusion approach is characterized by a maximal interaction strength in
the pseudo-critical region~\cite{Riek:2010fk,He:2011yi}, its application at
lower energies can help to test the validity of this behavior.

Our paper is organized as follows. In Sec.~\ref{sec_trans}, we briefly review
the ingredients to our model, \ie, the microscopically calculated HF transport
coefficients and fits of the macroscopic hydro evolution to bulk-hadron
observables at 62.4\,GeV (Sec.~\ref{ssec_trans}), and discuss in some detail
our implementation of the Cronin effect (Sec.~\ref{ssec_cronin}).
In Sec.~\ref{sec_obs}, we analyze the interplay of Cronin effect and HF
thermalization, including the partitioning of charm and bottom contributions
to HF electrons (Sec.~\ref{ssec_obs1}), and discuss our final numerical
results for their $R_{\rm AA}$ and $v_2$ in the context
of recent PHENIX and STAR data (Sec.~\ref{ssec_obs2}). We summarize
and conclude in Sec.~\ref{sec_concl}.

\section{Heavy-Flavor Transport and Initial Conditions}
\label{sec_trans}

\subsection{Transport Coefficients and Bulk Medium}
\label{ssec_trans}
Our nonperturbative transport model for open heavy flavor in URHICs was
introduced in Ref.~\cite{He:2011qa}. A strong-coupling approach is realized
in terms of both micro- and macro-physics, with nonperturbative scattering
amplitudes for HF interactions in the QGP and hadronic matter, and a
hydrodynamic medium evolution, respectively.  In the QGP, HQ interactions
with surrounding quarks, anti-quarks and gluons are evaluated using
in-medium $T$-matrix interactions~\cite{Riek:2010fk,Huggins:2012dj} based
on potentials motivated by thermal lattice-QCD (lQCD) results for the HQ
internal energy (including relativistic corrections). The pertinent
interaction strength increases as $T_{\rm pc}\simeq 170$\,MeV is approached
from above, inducing resonance correlations, which are subsequently utilized
as a hadronization mechanism in the resonance recombination model
(RRM)~\cite{Ravagli:2007xx} through a hydrodynamic hypersurface; left-over
charm and bottom quarks are hadronized via FONLL
fragmentation~\cite{Braaten:1994bz,Kartvelishvili:1977pi}, consistent with
the initial spectra in $pp$. The diffusion of $D$ and $B$ mesons is seamlessly
continued in the hadronic phase using effective scattering amplitudes off bulk
hadrons ($\pi$, $K$, $\eta$, $\rho$, $\omega$, $K^*$, $N$, $\bar N$, $\Delta$
and $\bar\Delta$). The hydrodynamic evolution is based on the 2+1D ideal hydro
code AZHYDRO~\cite{Kolb:2003dz}, augmented with a modern lQCD equation
of state for the QGP which is matched in a near-smooth transition to a hadron
resonance gas (HRG) at $T_{\rm pc}$=170\,MeV. The HRG is chemically frozen out
at $T_{\rm ch}$=160\,MeV. The introduction of a non-vanishing initial radial
flow, together with a compact initial entropy density profile, leads to a
saturation of the bulk-$v_2$ close to $T_{\rm pc}$ and mitigates the problems
of ideal hydrodynamics in describing the observed bulk-hadron $v_2$ down to
kinetic freezeout at $T_{\rm fo}$$\simeq$110\,MeV. It also appears to aid in the
understanding of direct-photon spectra and elliptic flow~\cite{vanHees:2014ida}.

In the following, we deploy our approach to study HF transport in Au+Au collisions
at $\sqrt{s_{\rm NN}}$=62.4\,GeV. The microscopic HF transport coefficients
in QGP and hadronic matter are assumed to be the same as at top RHIC and LHC
energies, since the quark chemical potential is still rather small at 62.4~GeV,
$\mu_q=\mu_B/3\simeq 20$\,MeV, and the coefficients are rather insensitive to
moderate variations in the quark/anti-quark composition (the contributions of
$Qq$ and $Q\bar q$ scattering are nearly identical, since the weaker interaction
strength in the resonant color-triplet relative to the color-singlet channel is
essentially compensated by the degeneracy in the
former~\cite{vanHees:2007me,Riek:2010fk}). The particle-antipartilce asymmetry is
further washed out in HF electron observables (which we compare to experiment),
as these typically involve an average over $e^+$ and $e^-$ (to improve statistics).

\begin{figure} [!t]
\includegraphics[width=1.05\columnwidth]{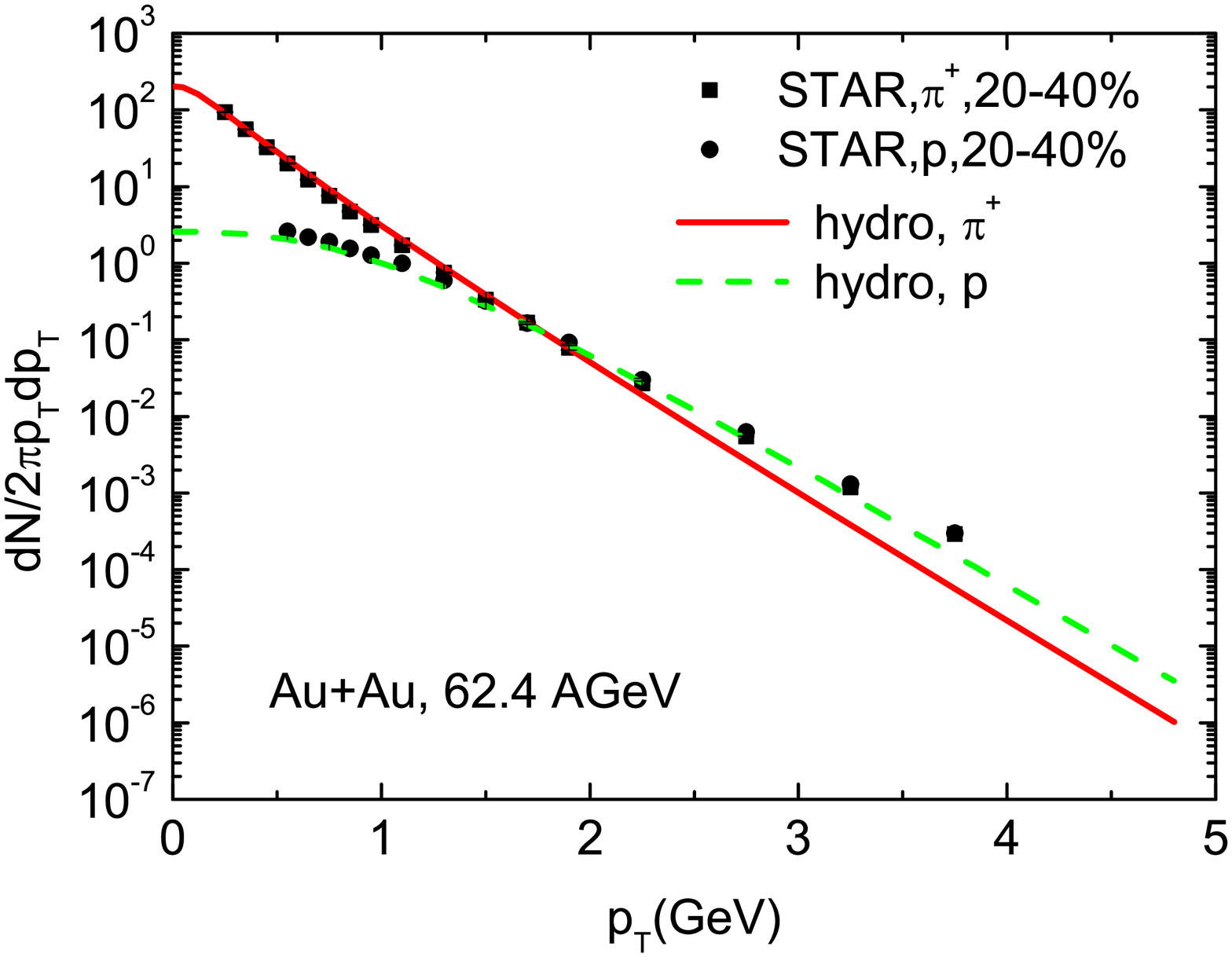}
\vspace{-0.5cm}
\includegraphics[width=1.05\columnwidth]{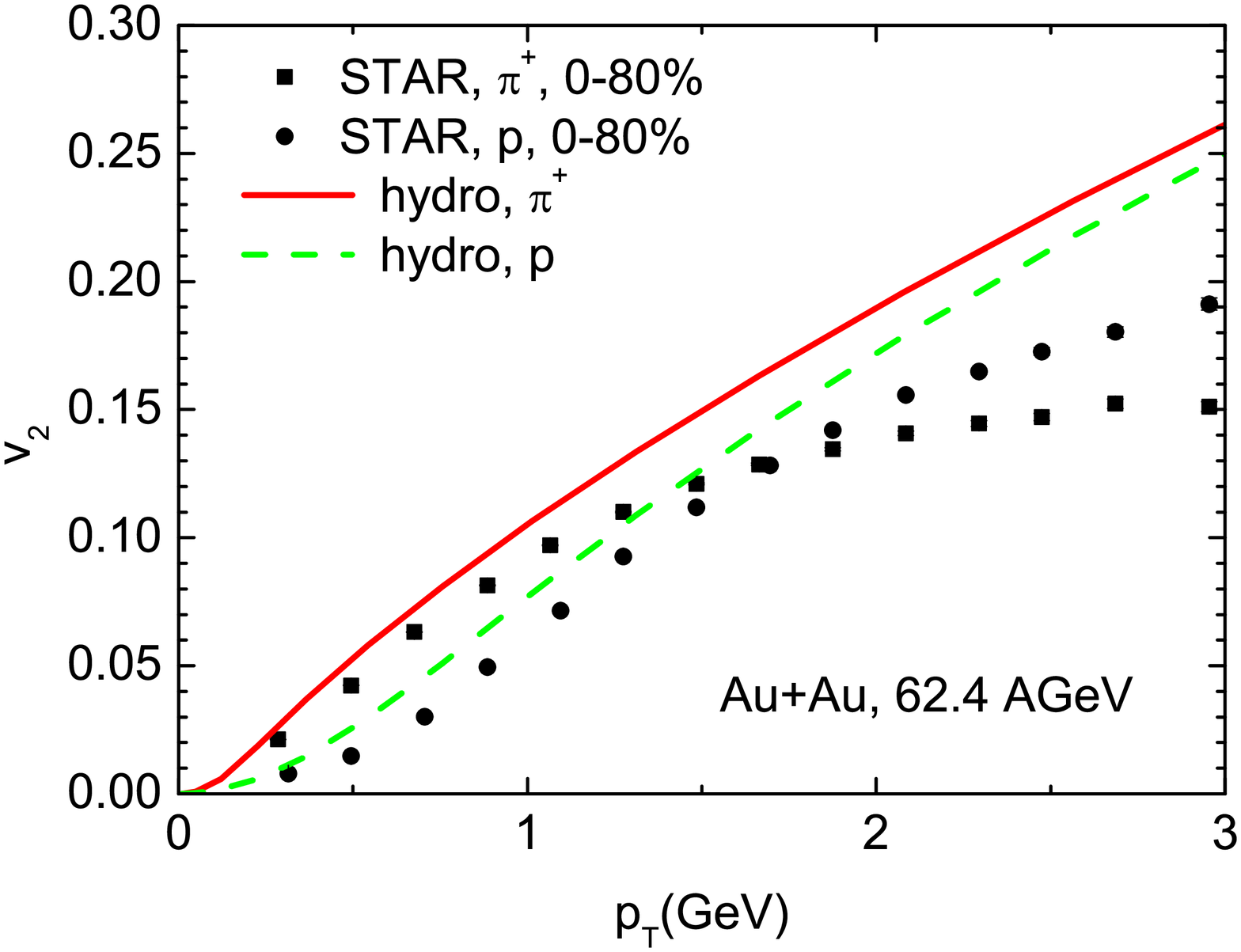}
\vspace{-0.3cm}
\caption{(Color online) Hydrodynamic fits of $\pi^+$ and proton spectra and $v_2$
to RHIC data~\cite{Abelev:2007ra,Adamczyk:2013gw} using our AZHYDRO tune.}
\label{fig_pipfits}
\end{figure}
The hydrodynamic bulk evolution is adapted as follows. For the equation of state
(EoS), we stay with our fit~\cite{He:2011zx} to the $\mu_q$=0 lQCD results
for the QGP part (corrections come in at order $(\mu_q/\pi T)^2$ which are at
the sub-percent level). We also assume $T_{\rm pc}$=170\,MeV and $T_{\rm ch}$=160\,MeV
to be unchanged (the latter is within errors of standard HRG fits to observed
particle ratios). However, we amend the HRG part of the EoS to accurately reflect
the change in hadro-chemistry at the lower energy. While this has no noticeable
impact on the thermodynamic state variables (which again deviate by less than
1\% from their 200\,GeV counterparts), the $\bar p/p$ ratio is significantly
affected (it decreases from ca.~0.75-0.8 at 200\,GeV to ca.~0.45-0.5 at 62.4\,GeV);
this figures noticeably into the absolute norm of our fits to the proton spectra.
For the initial conditions, we follow our successful tune at top RHIC
energy~\cite{He:2011zx} in that we assume the initial entropy density in the
transverse plane to be proportional to the binary collision density calculated
from the optical Glauber model, $s(x,y;\tau_0)=C(b)n_{\rm BC}(x,y;b)$.
Since the initial parton densities are smaller, we assume a slightly smaller
initial radial flow, $v_T(r;\tau_0)=\tanh(\alpha_0 r)$
with $\alpha_0=0.035~{\rm fm^{-1}}$, and later thermalization time,
$\tau_0=0.9~{\rm fm/c}$. When running our hydro with this initialization, we find
that a somewhat larger kinetic freezeout temperature of $T_{\rm kin}=130~{\rm MeV}$
yields a fair fit to pion and proton spectra out to $p_T\simeq$\,2-3\,GeV,
cf.~Fig.~\ref{fig_pipfits}. The resulting $v_2$ tends to slightly overestimate
the data already at lower momenta, possibly due to the lack of viscosity in our
hydro. However, we recall that the bulk-$v_2$ is mostly determined by the
low-momentum hadrons which constitute the major portion of the total yield; in
this regime our fit is not far off and thus should give a reasonable background
medium for HF diffusion, within an estimated 20\% of accuracy. Since our HF
transport coefficients fall off markedly with 3-momentum, most of their
interactions with bulk particles occur in the low-$p_T$ regime, where the
bulk-$v_2$ fit is rather close to the data.

\subsection{Initial HQ Spectra and Cronin Effect}
\label{ssec_cronin}
\begin{figure} [!t]
\includegraphics[width=1.05\columnwidth]{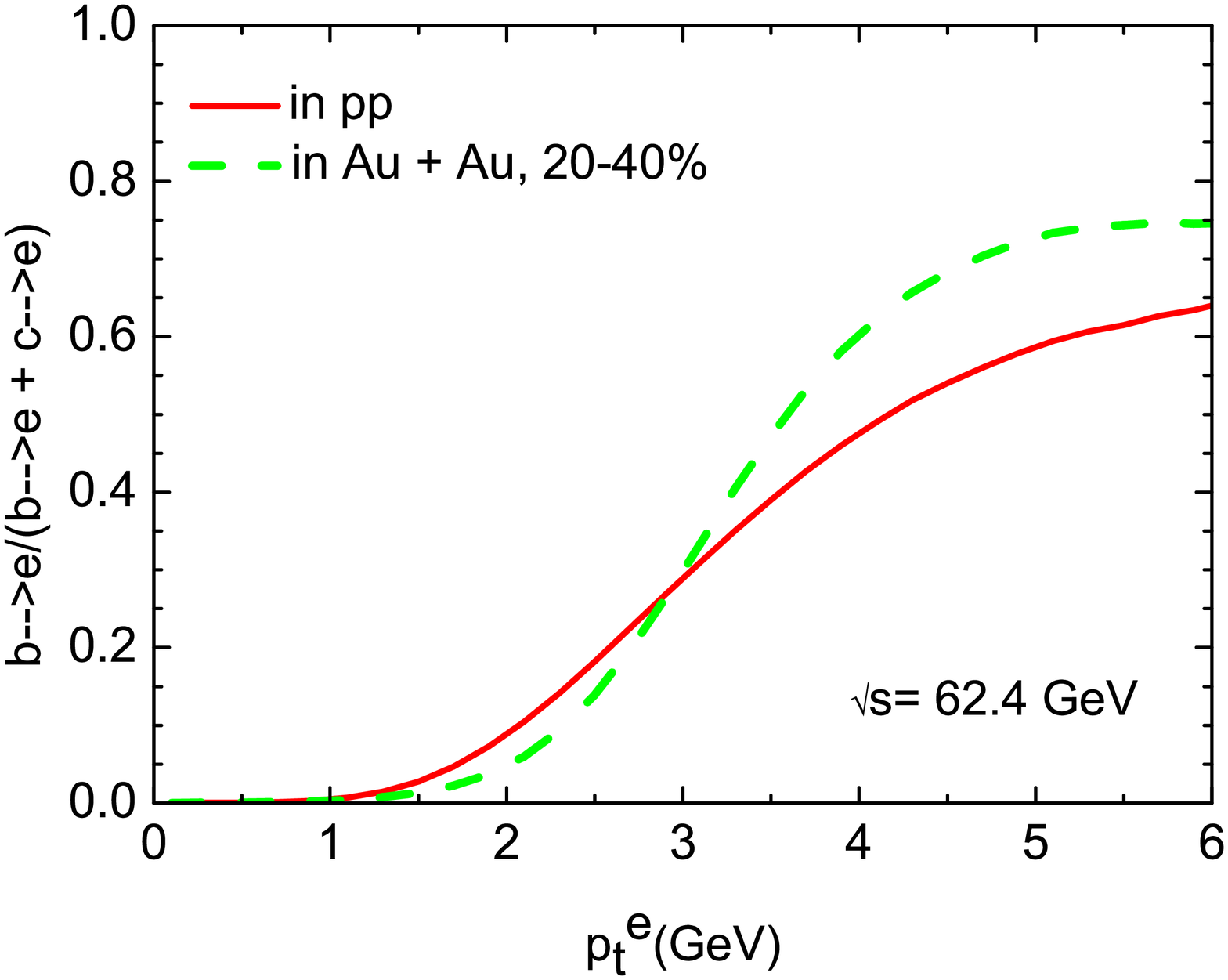}
\vspace{-0.5cm}
\caption{(Color online) Ratio of electrons from bottom decays to the sum of
charm + bottom in pp (red solid line) and 20-40\% central Au+Au collisions
(including medium effects, green dashed line). Only a slight change in the
shape of the $p_t^e$-dependent ratio is found for Au+Au.}
\label{fig_b-c-e}
\end{figure}

The spectra of $D$ and $B$ mesons in $pp$ collisions at
$\sqrt{s_{\rm NN}}=62.4~{\rm GeV}$ have not been measured yet. For the initial
conditions of the HQ spectra, we first generate them from the FONLL
software~\cite{Cacciari:2001td} followed by conversion into $D$- and $B$-meson spectra
using FONLL fragmentation functions~\cite{He:2014cla}. This also defines the
denominator of the pertinent nuclear modification factors, $R_{AA}$. The
resulting bottom-to-charm ($b/c$) cross section ratio of $1.9\times10^{-3}$ is
significantly (much) smaller than at top RHIC (LHC) energy,
$5\times10^{-3}$($5\times10^{-2}$)~\cite{He:2011qa,He:2014cla}. However,
when plotting the bottom fraction in the decay electron spectra vs.~electron
transverse momentum ($p_t^e$), the result is quite comparable to what we found
for top RHIC and LHC energies~\cite{He:2011qa,He:2014cla}, reaching 0.5 at
$p_t^e$=4-5\,GeV, cf.~Fig.~\ref{fig_b-c-e}. Thus, the steeper spectra
at lower energies decrease the inclusive $b/c$ fraction, while their
ratio as a function of $p_t^e$ seems to exhibit rather little dependence on
collision energy (a similar trend is found for the $B$-meson feeddown fraction
to $J/\psi$ production). This points at the importance of disentangling bottom
and charm contributions also at the lower energies.

Next, we turn to the cold-nuclear-matter (CNM) effects on the HQ spectra,
occuring prior to thermalization of the medium. These are usually associated
with nuclear shadowing of the parton distribution functions and a Cronin
broadening in $p_T$ due to scattering prior to the hard $Q\bar Q$ production
process. While shadowing of HQ production (especially for charm) presumably
plays a significant role in the interpretation of the HF spectra at LHC
energy~\cite{He:2014cla}, the Cronin effect is believed to become the most
important CNM effect toward lower energies~\cite{Vitev:2006bi,Vogt:2001nh};
for our $\sqrt{s_{\rm NN}}=62.4~{\rm GeV}$ calculations we thus focus on the
latter. To implement it we follow the standard procedure of smearing the
initial HQ spectra from $pp$ collisions with a Gaussian distribution
\begin{equation}
\label{Cronin}
g(\vec{k}_T)=\frac{1}{\pi \langle k_T^2\rangle}\exp(-{\vec{k}_T}^2/
\langle \Delta k_T^2 \rangle_{AA}) \
\end{equation}
with a variance parameter, $\langle \Delta k_T^2 \rangle_{AA}$, characterizing the nuclear
broadening. We estimate it following Ref.~\cite{Alberico:2011zy}, based on an
extension of what has been done before for
quarkonia~\cite{Hufner:2001tg,Vogt:2001nh}. Assuming the dominance of gluon
fusion processes, the nuclear broadening has been parameterized by the ansatz
\begin{equation}
\label{kt2}
\langle \Delta k_T^2\rangle_{\rm AA}^{Q\bar Q} = a_{gN} L_{\rm AB}(x,y;b) \ ,
\end{equation}
where $a_{gN}$ represents the transverse-momentum squared per path length of
a gluon traversing the nuclear medium, and $L_{\rm AB}(x,y;b)$ is the
sum of the path lengths of each gluon prior to producing the $Q\bar Q$ pair.
Typical values extracted from $p$A/$d$A spectra at top SPS~\cite{Topilskaya:2003iy}
and RHIC energy~\cite{Adare:2007gn} are $a_{gN}$=0.08\,GeV$^2$/fm and
$a_{gN}$=0.1-0.2\,GeV$^2$/fm, respectively~\cite{Zhao:2010nk}.
A more systematic energy dependence has been inferred in Ref.~\cite{Alberico:2011zy}
through the inelastic $NN$ cross section, $\sigma_{NN}(s)$, as
\begin{equation}
a_{gN}(s) =\Delta^2(\mu) \sigma_{NN}(s) \rho_0 \ ,
\end{equation}
where $\rho_0$ is the central nuclear density and
\begin{equation}
\Delta^2(\mu)=0.225\frac{\ln^2(\mu/{\rm GeV})}
{1+{\ln}(\mu/{\rm GeV})}{\rm GeV^2} \ ,
\end{equation}
characterizes the square-momentum transfer on the gluon per $NN$
collision~\cite{Wang:1998hs}; assuming its scale dependence given by the
HQ mass, $\mu\simeq 2m_Q$, results in approximate agreement with empirical
values for $a_{gN}$ at SPS and RHIC. Applying this framework to single HQ
spectra yields about half the broadening as for quarkonia,
$\langle \Delta k_T^2\rangle_{\rm AA}^{Q} =
\langle \Delta k_T^2\rangle_{\rm AA}^{Q\bar Q} /2$~\cite{Alberico:2011zy}.
Note that Eq.~(\ref{kt2}) keeps track of the broadening dependence on transverse
position and impact parameter through the effective length $L_{\rm AA}(x,y;b)$
(its explicit form can be found, \eg, in
Refs.~\cite{Hufner:2001tg,Zhao:2010nk,Alberico:2011zy}).
This automatically incorporates the binary collision scaling distribution of
heavy quarks in the transverse plane. Typical values for
$\langle \Delta k_T^2\rangle_{\rm AA}^Q$ in central Au-Au collisions at 62.4\,GeV
come out as $\sim$0.3(1.0)\,GeV$^2$ for charm (bottom) quarks. This results
in an enhancement of the HF $e^\pm$ spectra in d-Au of up to $\sim$15\%.
However, already in 200\,GeV $d$-Au an enhancement of up to $\sim$30-40\% is
observed~\cite{Adare:2012yxa}; for better agreement with phenomenology, we
therefore amplify the above estimates of the broadening parameters by a
factor of 2 (from here on referred to as default values), \ie,
$\langle \Delta k_T^2\rangle_{\rm AA}^{Q}\simeq 0.6~(2.0)~{\rm GeV^2}$
for charm (bottom) quarks in central Au-Au collisions. The resulting enhancement
in 200\,GeV $d$-Au collisions does not quite reach the observed $\sim$40\%, although
it is consistent within errors. It is conceivable that final-state effects
(\eg, coalescence~\cite{Hwa:2004zd}) provide an additional enhancement in these reactions (which we
do not further investigate here). To reflect the uncertainty in our procedure, we
vary the broadening parameters within $\pm$50\% of the default value.

\section{HF Electron Observables}
\label{sec_obs}
We are now in position to compute HF electron observables in 62.4\,GeV Au-Au
collisions, using relativistic Langevin simulations through QGP and hadronic
matter with the RRM hadronization interface, starting from FONLL HQ initial
spectra with Cronin broadening, and semileptonic final-state decays. We will
focus on the two standard observables, the nuclear modification factor,
\begin{equation}
\label{RAA}
R_{\rm AA}(p_T)=\frac{dN_{\rm AA}/dp_Tdy}{N_{\rm coll}dN_{\rm pp}/dp_Tdy} \ ,
\end{equation}
and the elliptic flow coefficient,
\begin{equation}
\label{v2}
v_2(p_T)=\left\langle \frac{p_x^2 - p_y^2}{p_x^2 + p_y^2} \right\rangle \ .
\end{equation}

\subsection{Cold vs. Hot Nuclear Medium Effects, and Charm vs. Bottom}
\label{ssec_obs1}
\begin{figure} [!t]
\includegraphics[width=1.05\columnwidth]{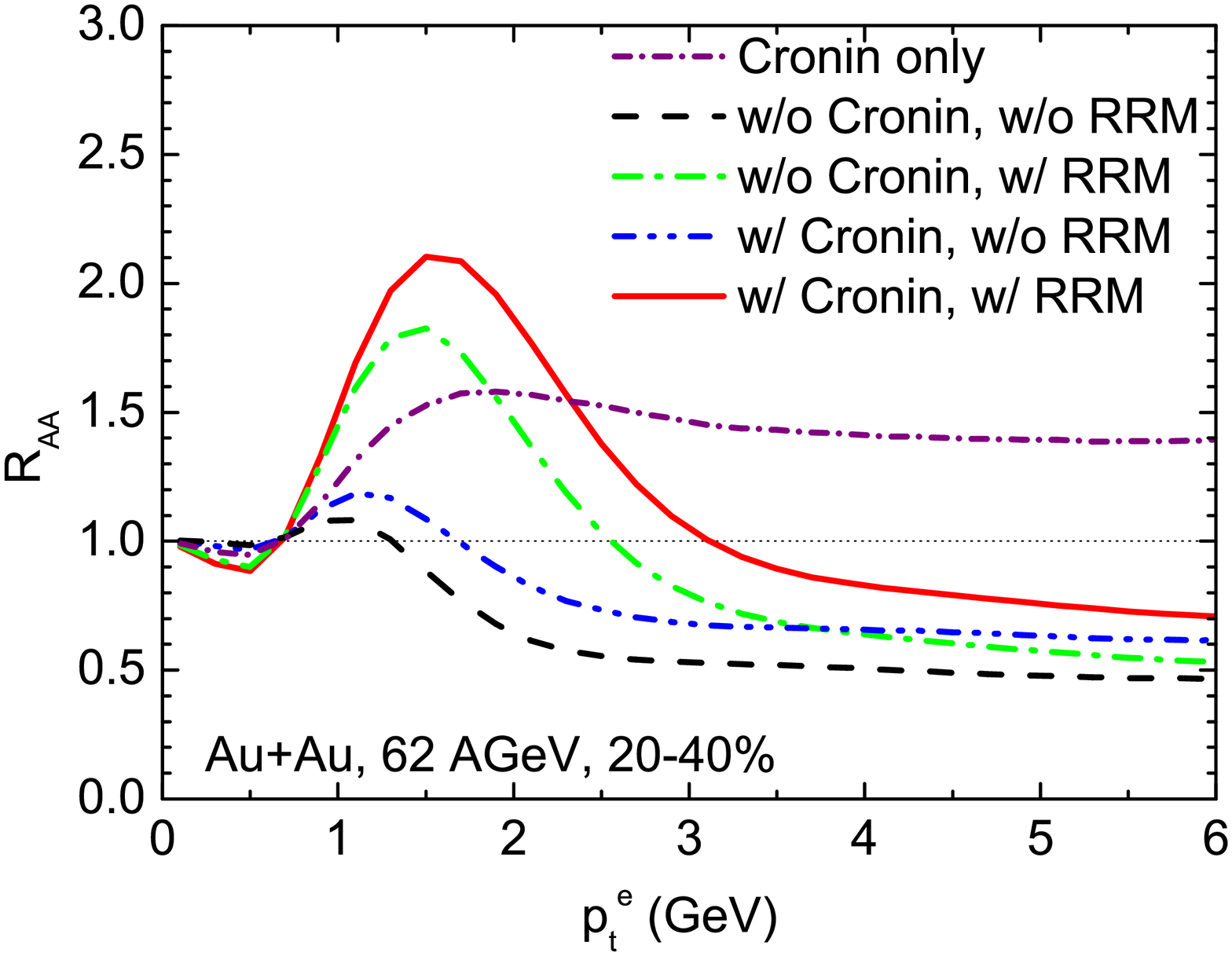}
\vspace{-0.5cm}
\caption{(Color online) Nuclear modification factor of HF electrons in
20-40\% central Au+Au collisions. Different scenarios are compared to illustrate
the effects of Cronin broadening and resonance recombination (applying the
default broadening and a $\sim$90\% integrated coalescence probability for
charm and bottom quarks).}
\label{fig_CroninvsRRM}
\end{figure}
Let us first study the magnitude and interplay of initial- and final-state
effects on the total (charm+bottom) $e^\pm$ $R_{\rm AA}$ in Au-Au collisions,
summarized in Fig.~\ref{fig_CroninvsRRM}. When only applying the Cronin broadening
at the default value (purple dash-dotted line), the enhancement factor reaches
$\sim$1.5 around $p_t^e\simeq2$\,GeV, slowly decreasing thereafter.
At $p_t^e\simeq$~5-6\,GeV, the enhancement is essentially due to the then
dominant bottom contribution.  Next, without any Cronin effect, we compare
the result of the HF diffusion with and without applying RRM as a hadronization
mechanism. This may serve as a lower estimate of the strong-coupling effects
around the pseudo-critical region. With neither RRM nor Cronin, a weak maximum
of the $R_{AA}$ above 1 develops around $p_t^e$$\simeq$1\,GeV (black dashed
line), reflecting the well-known ``flow bump" for heavy particles as a
consequence of the collectively expanding medium. Upon inclusion of RRM
(no Cronin; green dash-dotted line), the net addition of momenta from light
quarks from the thermalized medium generates a much more pronounced
``flow bump", shifted to a slightly larger $p_t^e$$\simeq$1.5\,GeV, with
a broader structure and a maximum value of $\sim$1.8; the onset of
suppression shifts up to above $\Delta p_t^e$$\simeq$2.5\,GeV.
When including the Cronin effect but without RRM (blue dash-dot-dotted
line), the maximum structure only moderately enhances over the diffusion-only
calculations (much less than due to RRM); however, toward high $p_T^e$ the
Cronin effect causes the $R_{AA}$ to level off at a higher value, by about 0.15
(or $\sim$30\%). Finally, when including both Cronin and RRM (red solid line),
the enhancement and broadening of the maximum structure of the electron
$R_{\rm AA}$ is further augmented, in a roughly additive manner relative to
the ``bare" Langevin baseline.
To summarize this study we find that the dominant effect of RRM (and thus,
in a way, of the strong HF coupling to the medium)  is the development of
a broad flow bump around $p_t^e$$\simeq$1.5\,GeV, while the Cronin effect
mainly manifests itself as an overall increase which is most significant
in the suppression regime at higher $p_t^e$.
The elliptic flow (not shown) is essentially unaffected by the Cronin effect,
but receives a roughly 30\% contribution from RRM, with $\sim$50\% from QGP and
$\sim$20\% from the hadronic phase~\cite{He:2012df}.

\begin{figure} [!t]
\includegraphics[width=1.05\columnwidth]{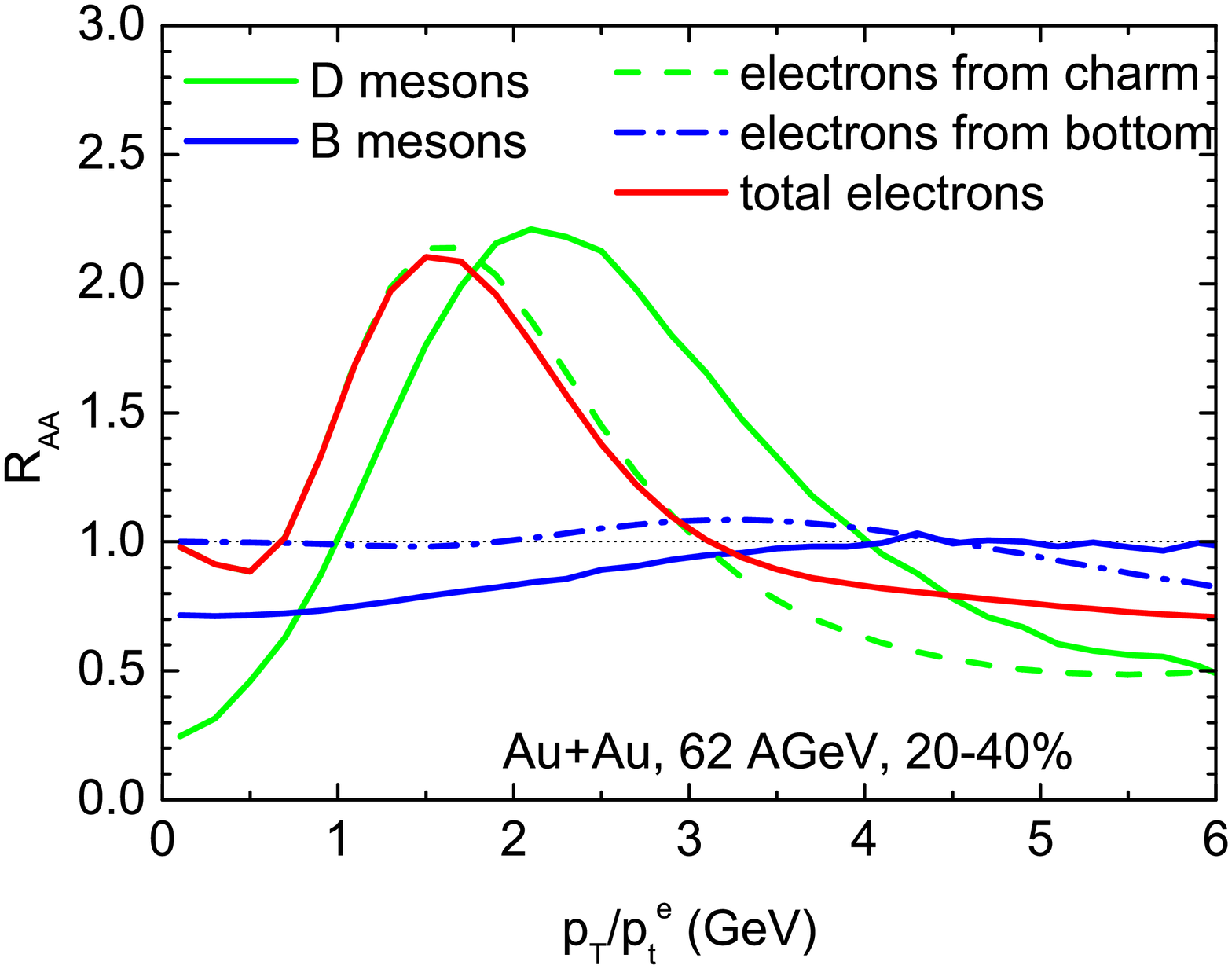}
\vspace{-0.5cm}

\includegraphics[width=1.05\columnwidth]{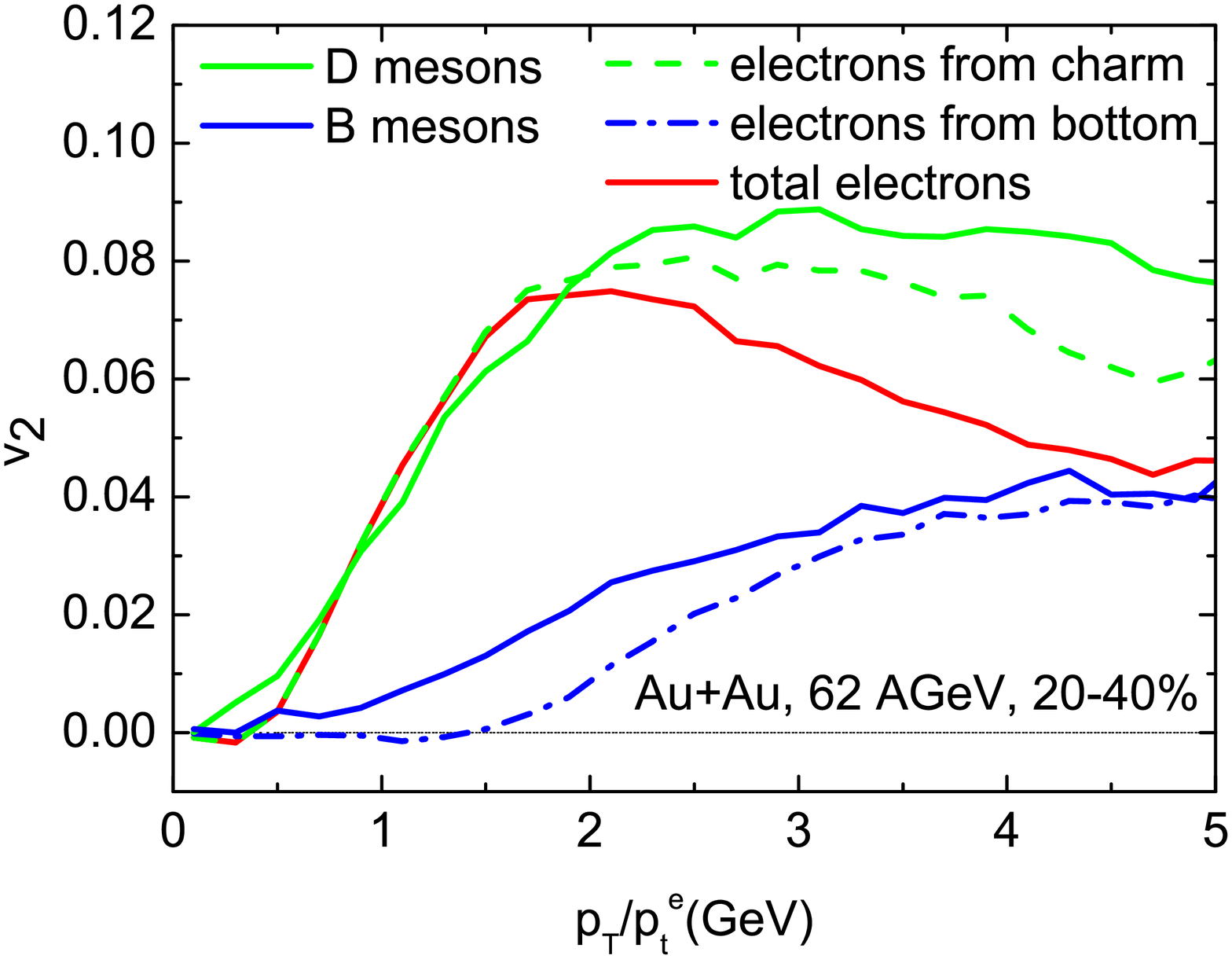}
\caption{(Color online) Upper panel: $R_{\rm AA}$ of $D$ mesons, $B$ mesons,
their individual and total $e^\pm$ decays in 20-40\% central Au+Au collisions.
Lower panel: the corresponding $v_2$ results. The default Cronin broadening and
$\sim$90\% integrated coalescence probability for charm/bottom quarks are applied.}
\label{fig_DevsBe}
\end{figure}
To disentangle the effects on the charm and bottom contributions, we show
our full results for $D$ and $B$-meson $R_{\rm AA}$ and $v_2$ and their
individual $e^\pm$ decay spectra in Fig.~\ref{fig_DevsBe}.  Due to the
factor-of-3 mass difference, the $D$- and $B$-meson $R_{\rm AA}$ and $v_2$
are quite different at a given $p_T$, which is essentially preserved at a given
(down-shifted) value of $p_t^e$. The total $e^\pm$ $R_{\rm AA}$ and $v_2$
largely follow the charm electrons up to $p_t^e\simeq$2-3\,GeV (due to
charm dominance), while for higher $p_t^e$ the bottom contribution causes
a increasing reduction in both suppression and $v_2$.

\subsection{Comparison to RHIC Data}
\label{ssec_obs2}
We now turn to a systematic comparison of our results to HF electron $R_{\rm AA}$
and $v_2$ data for different centralities~\cite{Adare:2014rly}. We assign
theoretical uncertainties due to the integrated coalescence probabilities
of charm and bottom quarks~\cite{He:2014cla} ($\sim$\,50-90\% as in our previous
work) and, additively, due to the Cronin effect (default broadening with
a $\pm$50\% variation), represented by shaded (colored) bands. As discussed in
Sec.~\ref{ssec_obs1}, uncertainties in the coalescence probability prevail in
the low $p_t^e$ region, while those from the Cronin effect take over at high
$p_t^e\simeq$\,5-6\,GeV.
\begin{figure} [!t]
\includegraphics[width=0.99\columnwidth]{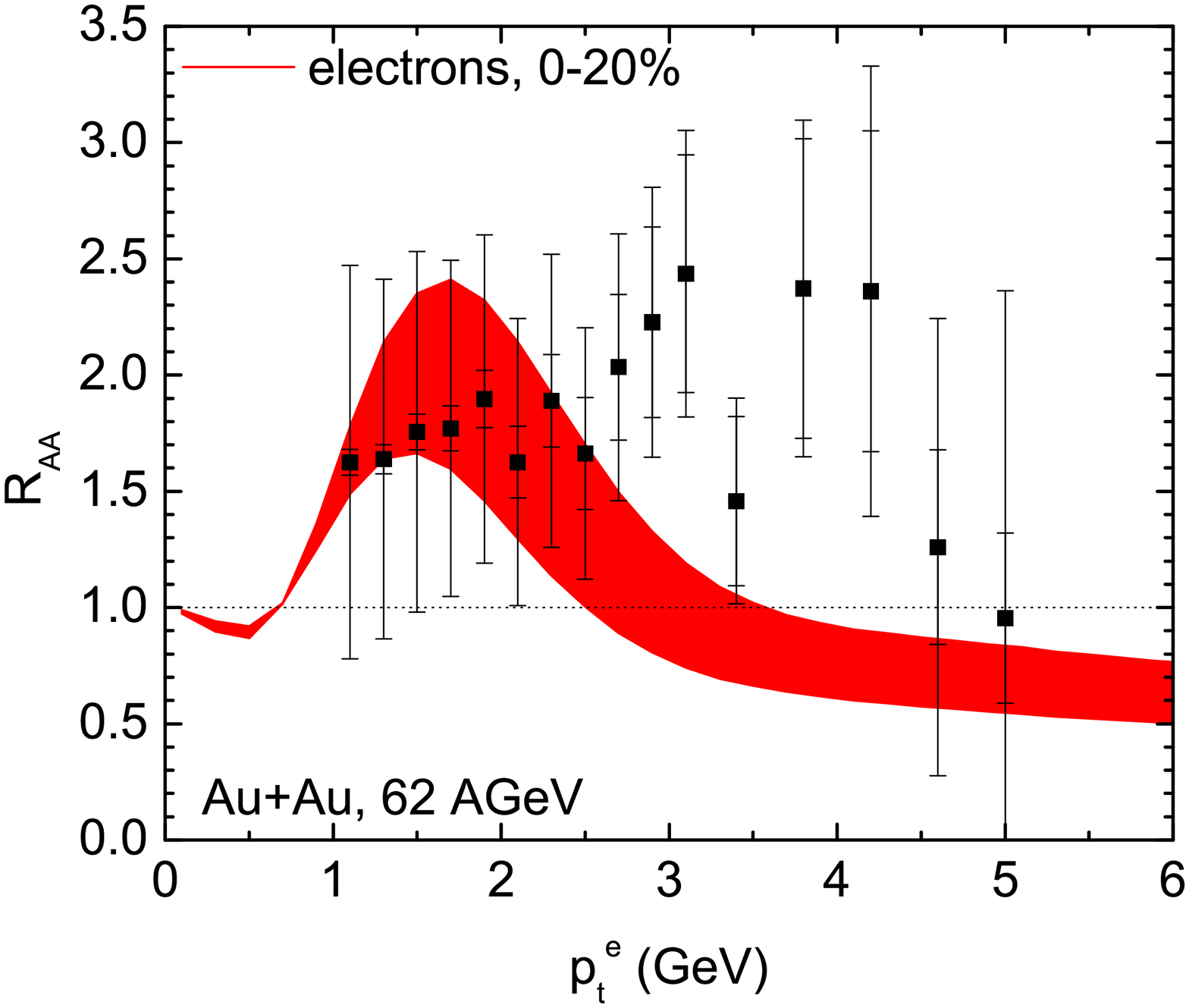}

\vspace{-0.15cm}

\includegraphics[width=0.99\columnwidth]{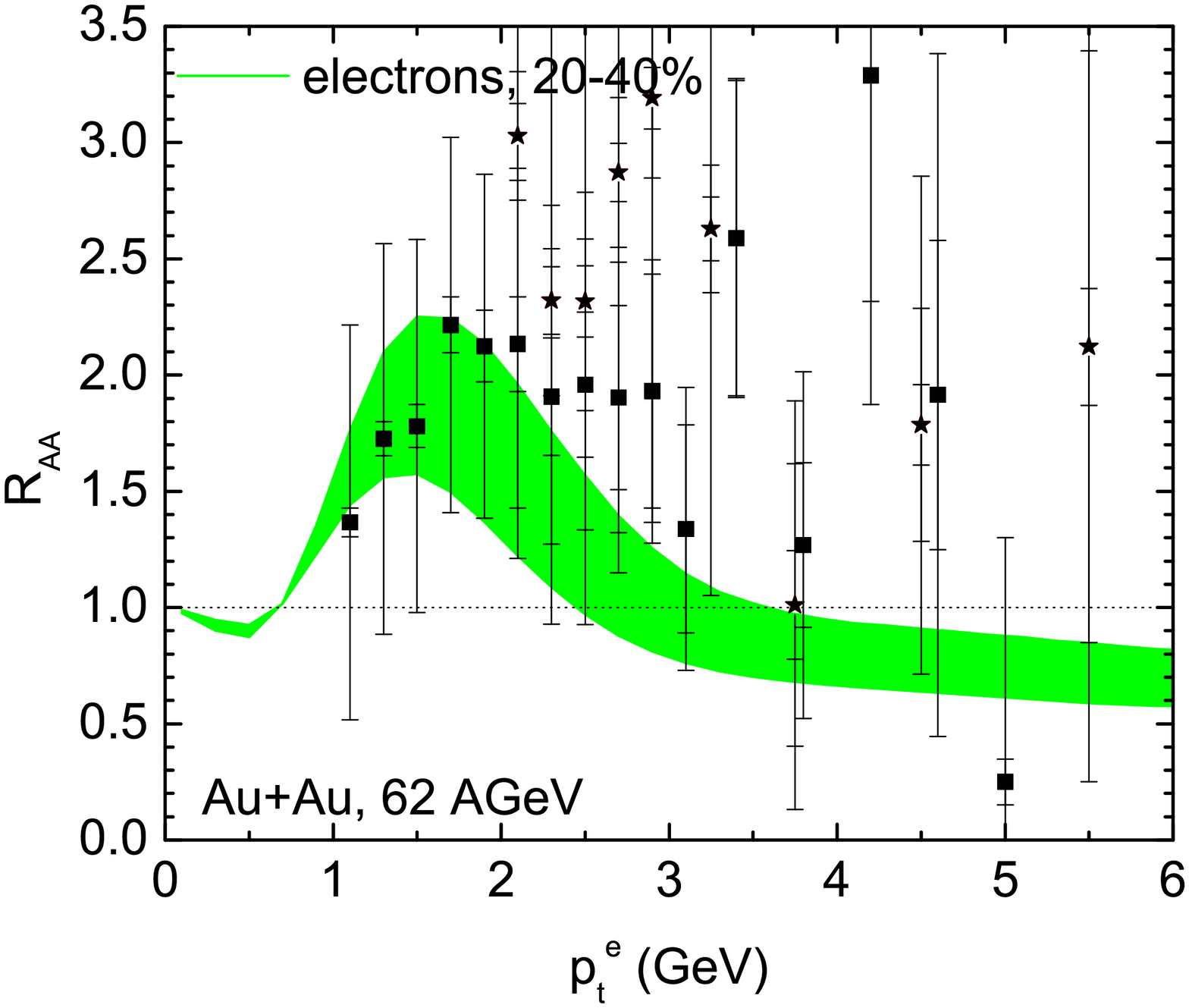}

\vspace{-0.15cm}

\includegraphics[width=0.99\columnwidth]{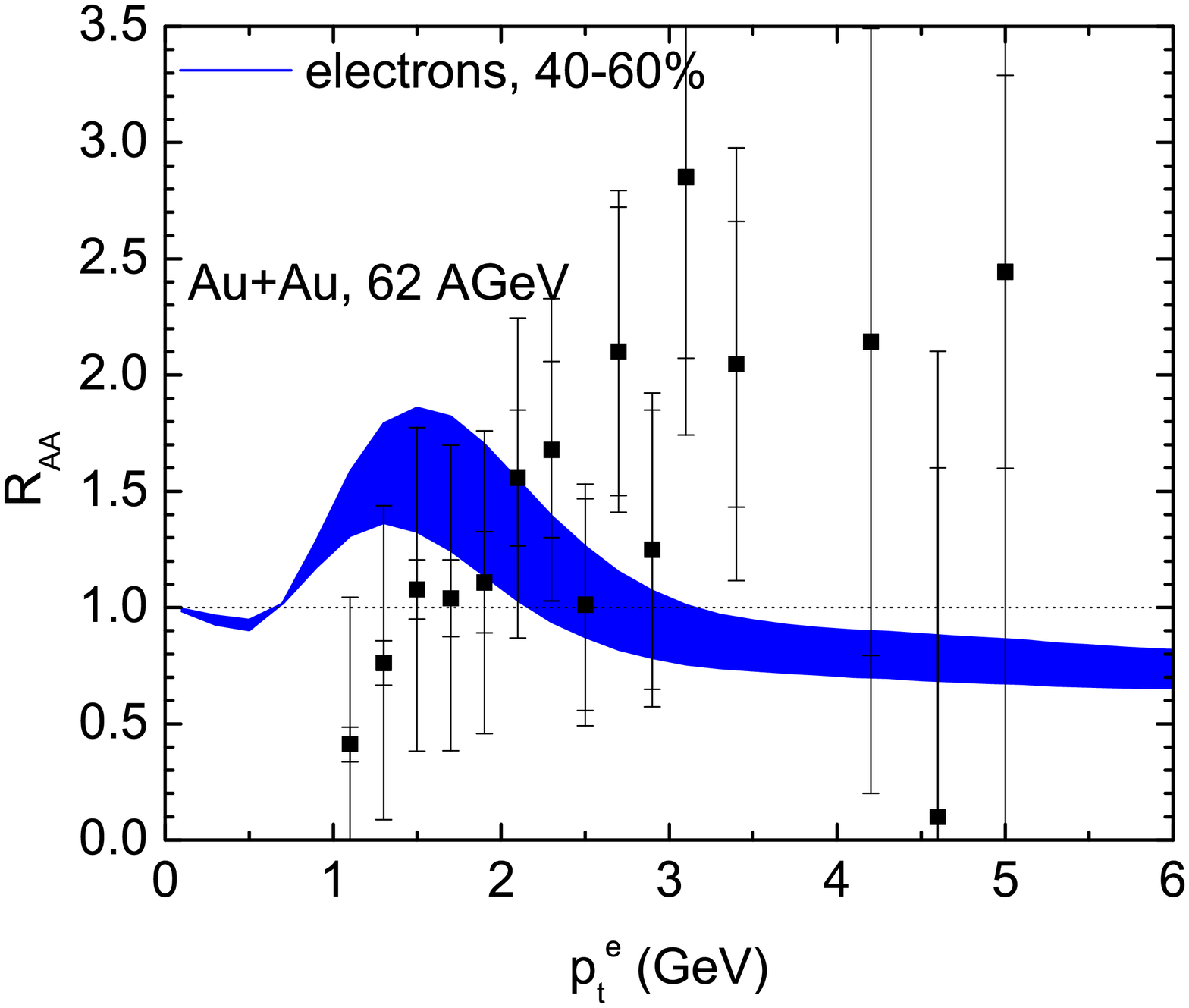}
\caption{(Color online) Total HF electron $R_{\rm AA}$ in
Au-Au($\sqrt{s_{\rm NN}}$=62.4\,GeV) collisions at 0-20\% (upper panel),
20-40\% (middle panel) and 40-60\% (lower panel) centrality. The bands
indicate uncertainties due to the total charm- and bottom-quark coalescence
probabilities of $\sim$50-90\% and the variation in the Cronin broadening.
PHENIX data (filled boxes) are taken from Ref.~\cite{Adare:2014rly}, preliminary
STAR data (filled stars in the middle panel) from Ref.~\cite{Mustafa:2012jh}.}
\label{fig_electrons_RAA}
\end{figure}


\begin{figure} [!t]
\includegraphics[width=1.0\columnwidth]{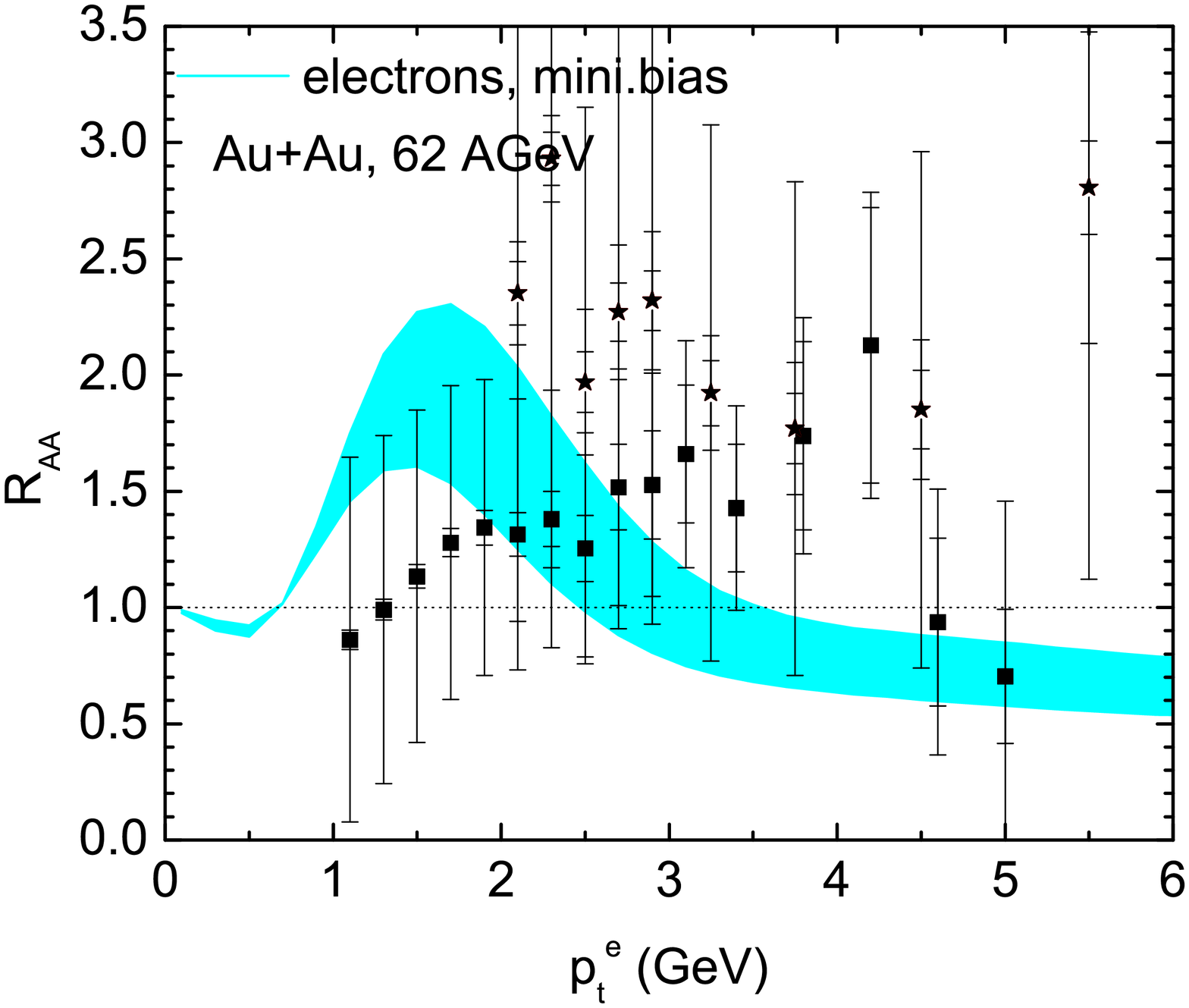}

\vspace{-0.15cm}

\includegraphics[width=1.0\columnwidth]{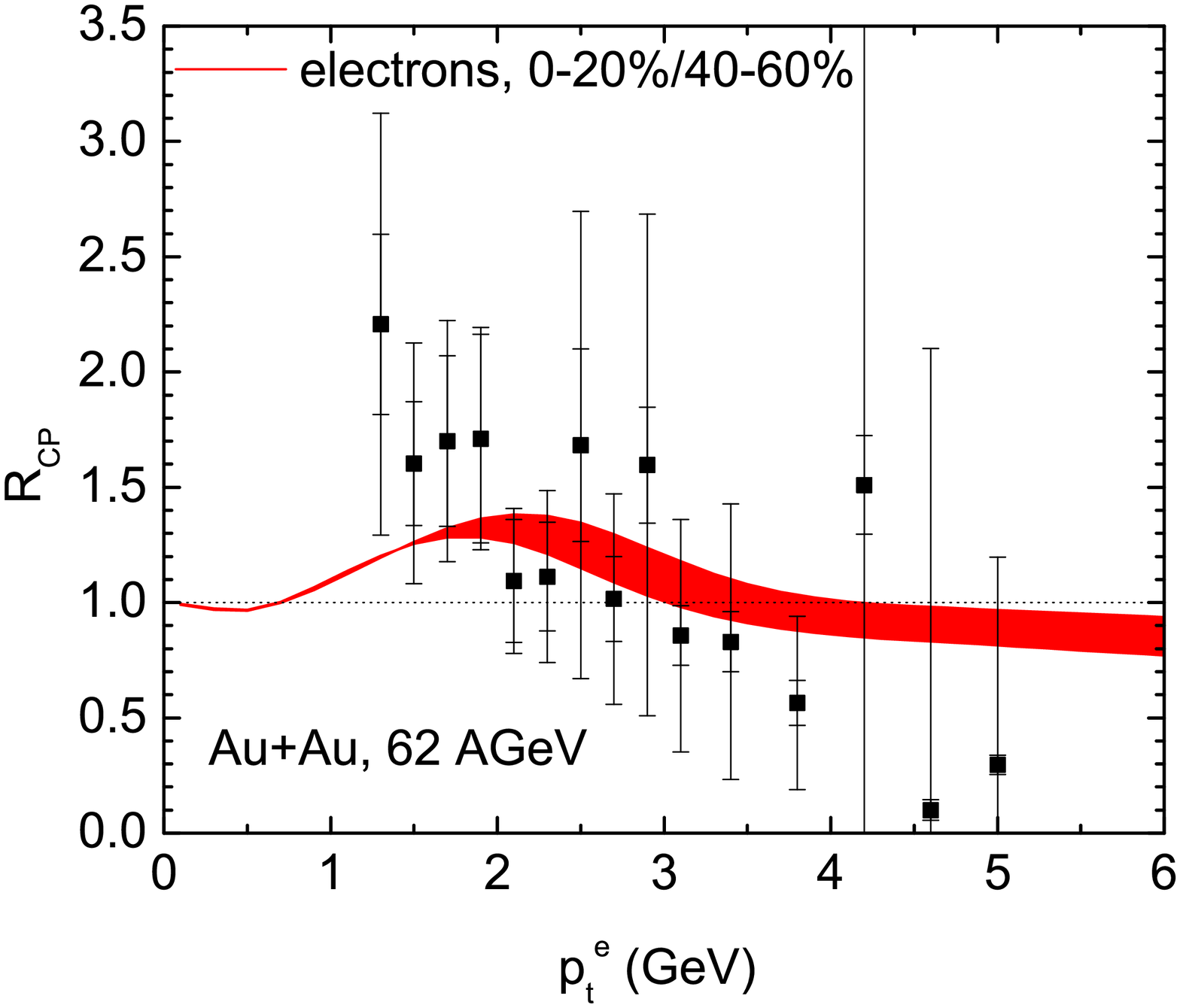}
\caption{(Color online) Total HF electron $R_{\rm AA}$ in minimum bias (upper panel)
Au-Au($\sqrt{s_{\rm NN}}$=62.4\,GeV), calculated from a $N_{\rm coll}$-weighted
average of the 0-20\%, 20-40\%, 40-60\% and 60-80\% bins, and $R_{\rm CP}$
(lower panel) between 0-20\% and 40-60\% centrality bins. PHENIX data (filled boxes)
are taken from Ref.~\cite{Adare:2014rly}, preliminary STAR data (filled stars in
the upper panel) from Ref.~\cite{Mustafa:2012jh}.}
\label{fig_electrons_RAA_minibias}
\end{figure}


\begin{figure} [!t]
\includegraphics[width=1.0\columnwidth]{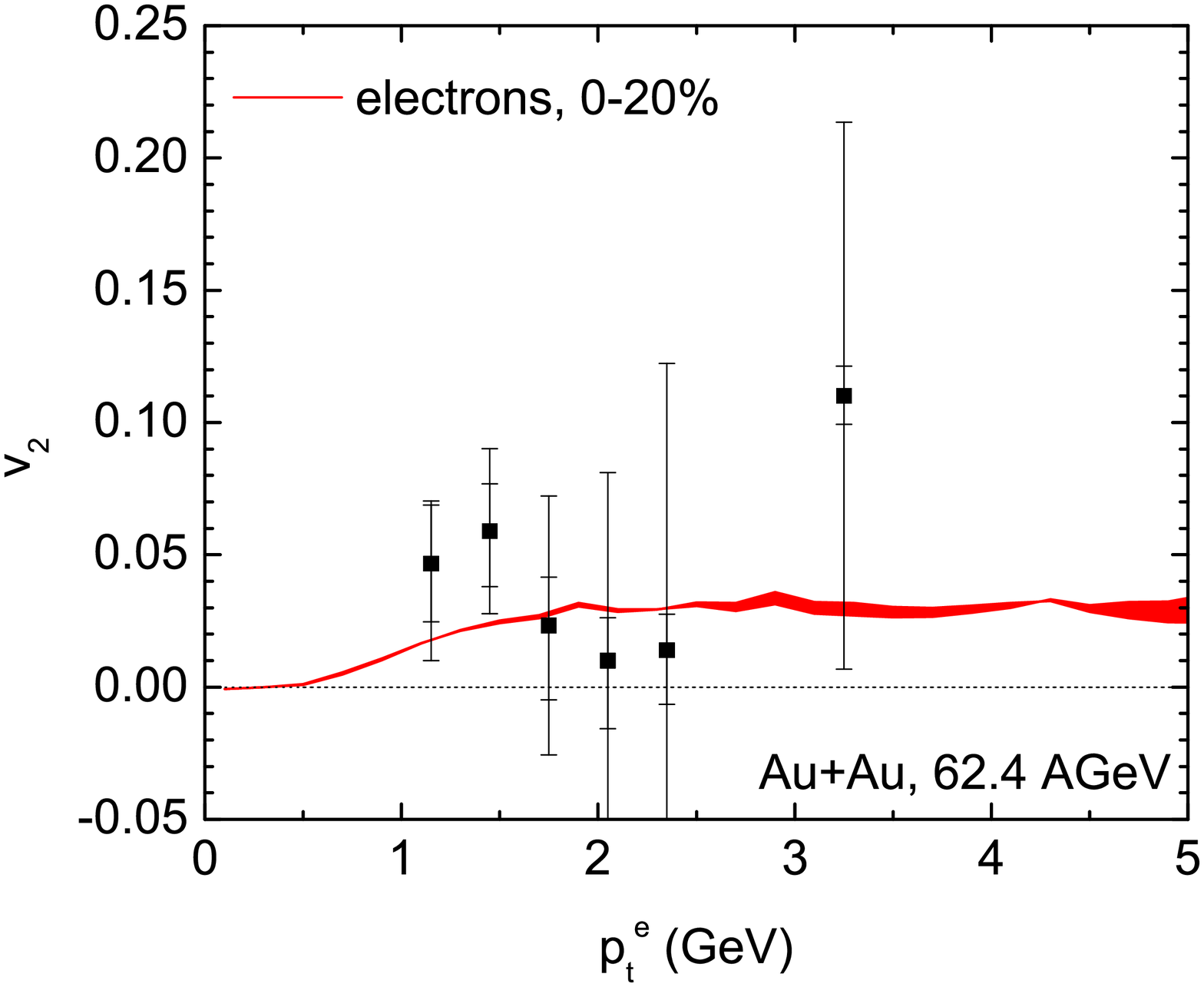}

\vspace{-0.15cm}

\includegraphics[width=1.0\columnwidth]{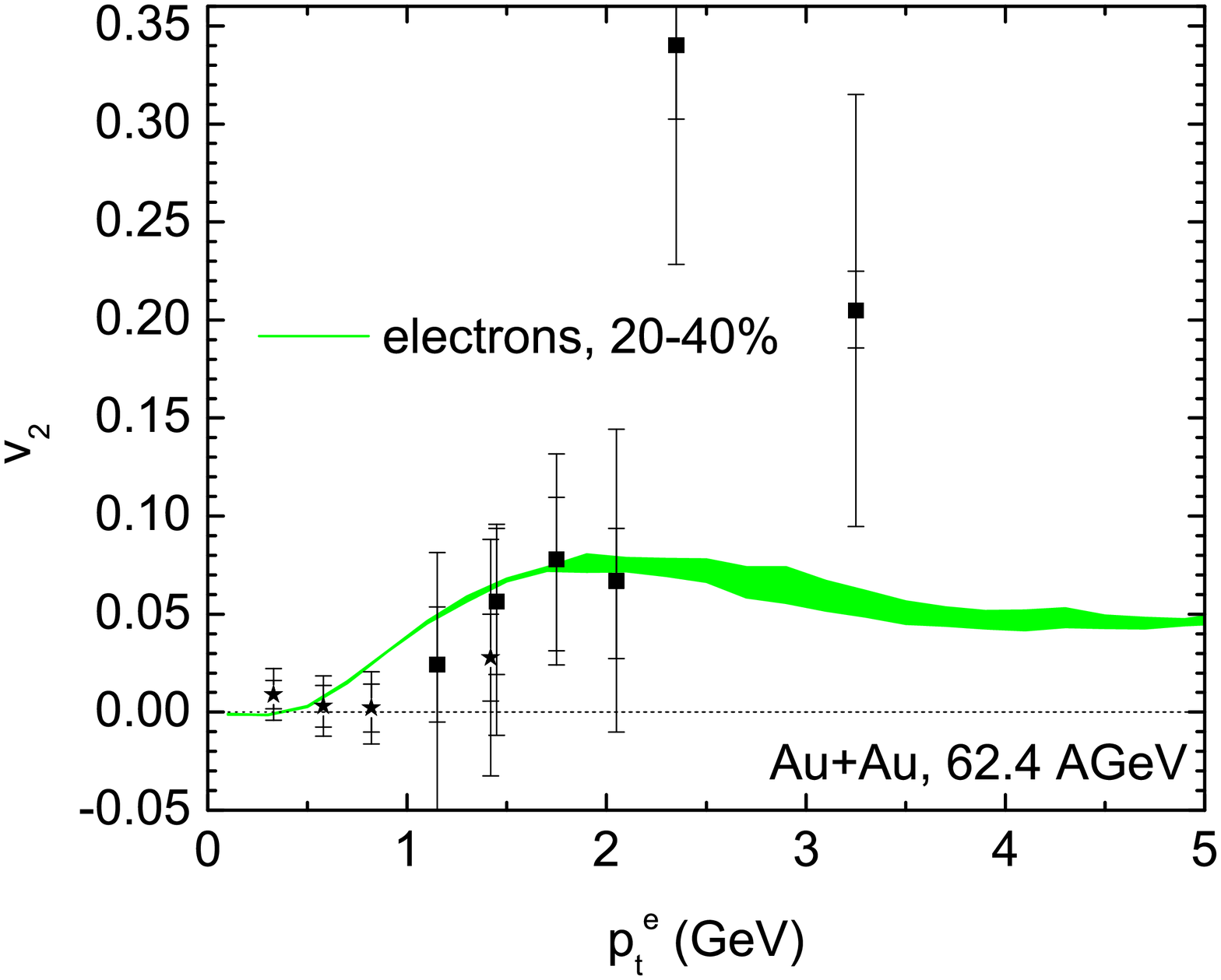}

\vspace{-0.15cm}

\includegraphics[width=1.0\columnwidth]{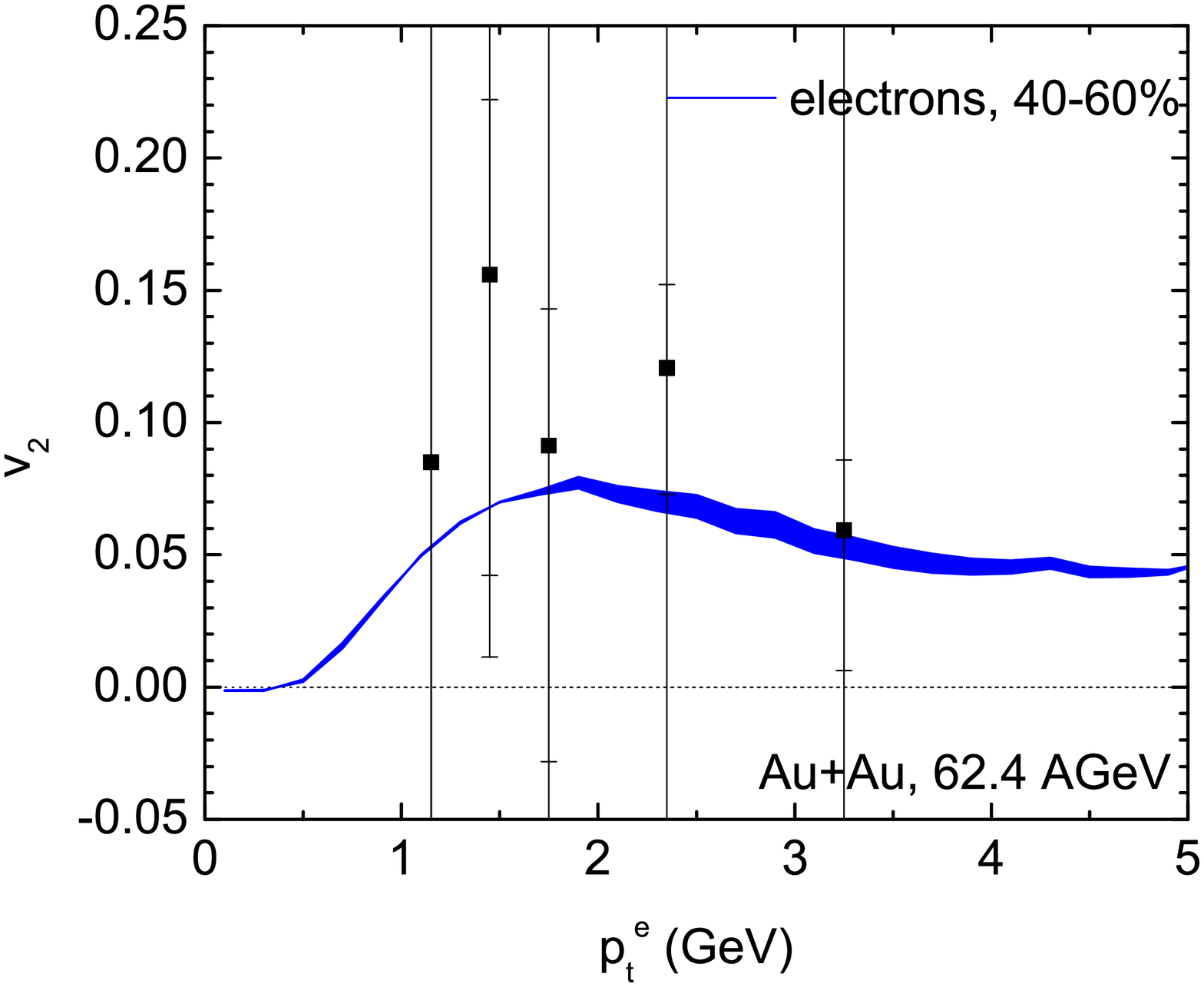}
\caption{(Color online) Total HF electron $v_2$ in 0-20\%,  20-40\%, 40-60\% central
Au-Au($\sqrt{s_{\rm NN}}$=62.4\,GeV) in the top, middle and lower panel, respectively.
The bands denote again the uncertainty in the integrated charm- and bottom-quark coalescence
probabilities of $\sim$50-90\%. PHENIX data (filled boxes) are taken from Ref.~\cite{Adare:2014rly}.}
\label{fig_electrons_v2}
\end{figure}

\begin{figure} [!t]
\includegraphics[width=1.0\columnwidth]{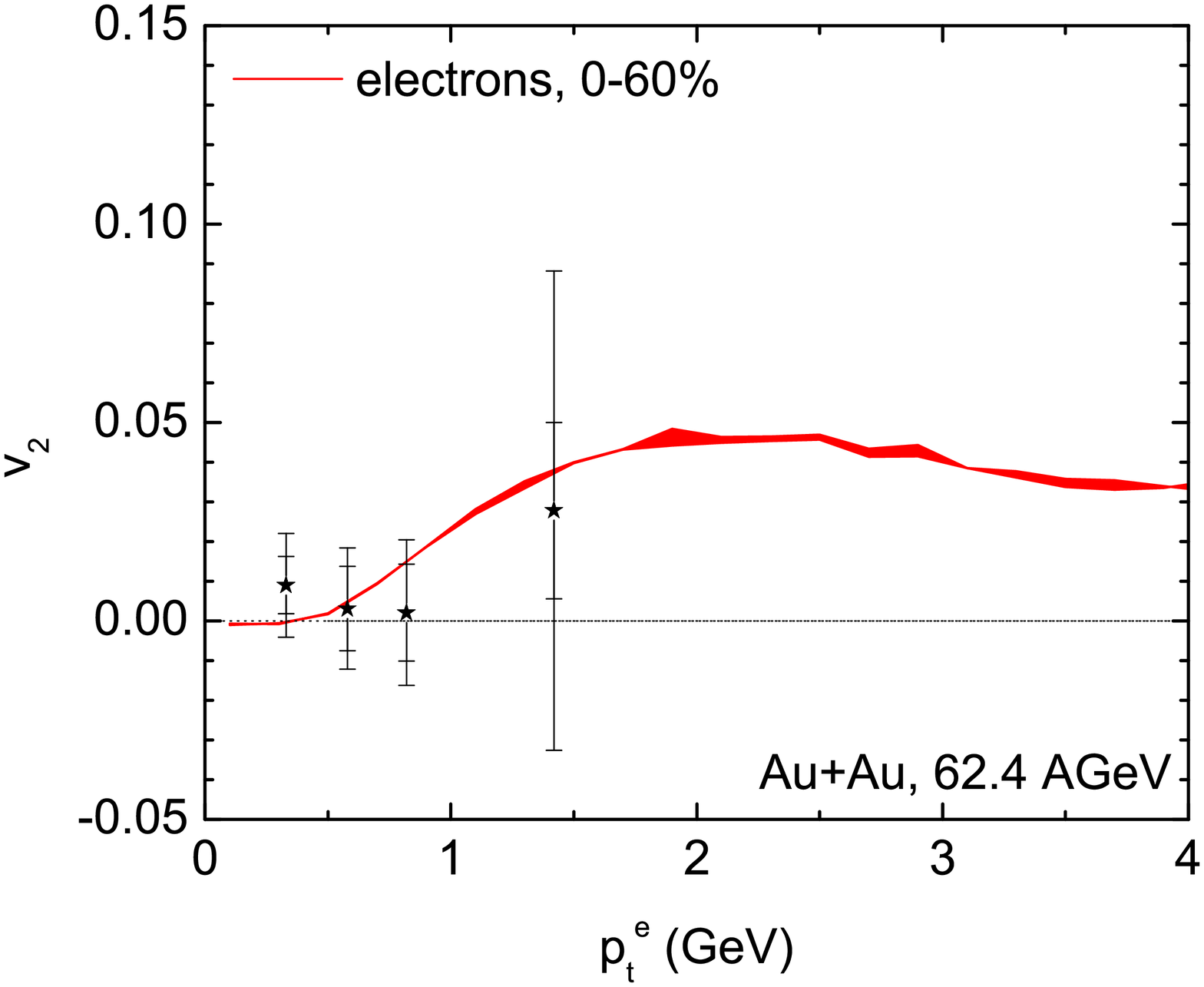}
\caption{(Color online) Total HF electron $v_2$ in 0-60\% central Au-Au($\sqrt{s_{\rm NN}}$=62.4\,GeV), calculated from a $N_{\rm coll}$-weighted
average of $v_2$'s of the 0-20\%, 20-40\% and 40-60\%  bins. STAR data (filled stars) are taken from Ref.~\cite{Adamczyk:2014yew}.}
\label{fig_electrons_v2_minibias}
\end{figure}

For the HF $R_{\rm AA}$'s, plotted in Fig.~\ref{fig_electrons_RAA}, our calculations
roughly reproduce the data in the two central bins (0-20\% and 20-40\%), with a tendency
to underpredict the yields at higher $p_t^e$. Given the current experimental errors,
it is inconclusive whether the data support a pronounced flow+Cronin bump as present
in the calculations. For the more peripheral bin (40-60\%) our calculation tends to
overestimate the data at low $p_t^e\le$~1\,GeV, where the measured decrease sets in
earlier than captured by our implementation of the Cronin effect and/or the collective
effects due to the coupling to the hot expanding medium (uncertainties in the $pp$
basline, or pre-equilibrium effects~\cite{Chesler:2013cqa}, which we do not consider
here, may also play a role). Since our previous results for the $D$-meson $R_{\rm AA}$
at top RHIC energy produce a flow bump which tends to peak at somewhat higher $p_T$
than STAR data in central collisions~\cite{Adamczyk:2014uip}, we believe that an
increased collectivity (to deplete the low-$p_t^e$ region) at 62.4\,GeV is not a
likely mechanism to resolve the discrepancy in the 40-60\% centrality bin. For
minimum bias (MB) collisions, displayed in the upper panel of
Fig.~\ref{fig_electrons_RAA_minibias}, we obtain our result by averaging the above
three plus the 60-80\% centrality bin, weighted by the binary-collision number,
$N_{\rm coll}$, of each bin.
The overprediction in the low-$p_t^e$ region found for 40-60\% centrality (and probably
also for 60-80\%) is somewhat mitigated in the MB sample but still present.

We have furthermore computed the central-to-peripheral ratio, $R_{\rm CP}$, obtained from
the ratio between the $R_{\rm AA}$'s of 0-20\% and 40-60\%, and compare it to PHENIX data
in the lower panel of Fig.~\ref{fig_electrons_RAA_minibias}. The $R_{\rm CP}$ has the
advantage that uncertainties due to the $pp$ baseline spectra largely drop out, but
the centrality dependence of the Cronin effect still affects it, although to a lesser
extent than in the $R_{\rm AA}$. Our calculated $R_{\rm CP}$ exhibits a significant
enhancement above 1 for $p_t^e=1\sim 3~{\rm GeV}$, as a genuine signature of the
stronger flow bump in the $R_{\rm AA}$ for central relative to peripheral collisions.
At the same time, and as a necessary consequence, the stronger suppression in more
central collisions still manifests itself at higher $p_t^e$$>$5\,GeV, although
the quenching is somewhat counteracted by the Cronin effect.
The current PHENIX data for $R_{\rm CP}$ support this trend, albeit again with large
uncertainties.

Finally we compare our calculated HF electron $v_2$ with PHENIX data for 3 centralities (0-20\%, 20-40\% and 40-60\%) in
Fig.~\ref{fig_electrons_v2}, and with STAR data for 0-60\% centrality in Fig.~\ref{fig_electrons_v2_minibias}. The latter
is calculated from a $N_{\rm coll}$-weighted
average of $v_2$'s of the former 3 centrality bins. Here, no discrepancy with
the data can be made out, albeit within rather large experimental uncertainties
at this point. It would be illuminating to scrutinize the agreement with higher
precision data. As emphasized in our previous works~\cite{vanHees:2007me,Rapp:2008zq},
the maximum interaction strength
around $T_{\rm pc}$ in our microscopic $T$-matrix model for HF diffusion, together with
a build-up time of the bulk-$v_2$ of a few fm/$c$, implies the HF $v_2$ to be more
sensitive to the pseudo-critical region than the $R_{\rm AA}$.
We expect this sensitivity to be enhanced at lower collision energies (but the latter
should still be high enough for the medium evolution to comfortably encompass the
transition region).

\section{Summary}
\label{sec_concl}
We have conducted a study of open HF probes in Au+Au collisions at
$\sqrt{s_{\rm NN}}$=62.4\,GeV using a nonperturbative transport model in a
hydrodynamically expanding medium. Heavy-flavor diffusion is realized within the
same transport approach through QGP, hadronization and hadronic matter as applied
in our previous calculations at top RHIC and LHC energies~\cite{He:2012df,He:2014cla}.
To the extent that the initial HQ distributions and hydrodynamic medium evolution
can be controlled, our results carry predictive power.
For the medium evolution, our earlier ideal-hydro tune at full RHIC energy (utilizing
an initial flow and compact entropy density profile) was adapted to reproduce measured
pion and proton spectra and $v_2$ at 62.4\,GeV, with acceptable precision for our
HF estimates. For the initial-state Cronin effect, we took guidance from theoretical
expectations with numerical values motivated by experiments at full RHIC energy.
The suppression and flow pattern of the HF decay electrons at 62.4\,GeV emerges
from an interplay of the Cronin enhancement and the partial thermalization of
heavy flavor in the hot medium starting from a softer $pp$ baseline than at 200\,GeV.
In particular, the role of resonance recombination, characterizing the strong HF-medium
coupling around the pseudo-critical region, has been identified as the main source
for developing a rather pronounced ``flow bump" in the electron $R_{\rm AA}$. Within
the currently rather large experimental uncertainties, our predictions are roughly
compatible with the PHENIX data, with a tendency to underpredict the yields at high
$p_t^e$ in central collisions and at low $p_t^e$ in peripheral ones. Whether these
discprepancies are due to the transport treatment or initial-state effects remains
an open question. On the other hand, our results for the $R_{\rm CP}$, which is
less sensitive to uncertainties in the initial spectra from $pp$, agree with the
data reasonably well, and still exhibit a transition from a flow bump at low
$p_t^e$ to a net suppression at high $p_t^e$. No discrepancies were found with
the $v_2$ data, where our calculations yield values comparable to full RHIC energy.
Our analysis is not incompatible with the formation of a QCD medium in Au+Au
collisions at 62.4\,GeV, whose strong coupling around the pseudo-critical
region imparts substantial collectivity on HF particles through their nonperturbative
interactions. However, more precise data would allow for much more sensitive tests
of the pertinent structures emerging from our calculations. \\

\acknowledgments We are indebted to F.~Riek and K. Huggins for providing
the results for the HQ transport coefficients, to W. Xie and M.~Mustafa for providing
the STAR data, and to J. M. Durham and A. Lebedev for providing the PHENIX
data. This work was supported by the U.S.~National Science Foundation (NSF) through CAREER grant PHY-0847538
and grant PHY-1306359, by the A.-v.-Humboldt
Foundation, by the JET Collaboration and DOE grant DE-FG02-10ER41682,
and by NSFC grant 11305089.

\end{document}